\documentclass[11pt,letterpaper,titlepage]{article}

\usepackage{times}
\usepackage{amsmath}
\usepackage{amsfonts}
\usepackage{amssymb}
\usepackage{hyperref}
\usepackage[usenames,dvipsnames]{xcolor}
\usepackage{algorithm}
\usepackage[noend]{algpseudocode}

\definecolor{light-gray}{gray}{0.75}
\algrenewcommand{\algorithmiccomment}[1]{\hskip3em{{\footnotesize \textcolor{light-gray}{$\blacktriangleright$}}} #1}

\usepackage{ifdraft}

\usepackage[pdftex]{graphicx}
\graphicspath{./figures/}
\DeclareGraphicsExtensions{.pdf}

\usepackage{float}
\usepackage{caption}
\usepackage[tight,footnotesize]{subfigure}

\ifoptionfinal{
\newcommand{\ysnote}[1]{}
\newcommand{\aknote}[1]{}
\newcommand{\cwnote}[1]{}
} {
\newcommand{\ysnote}[1]{  {\textcolor{magenta}      { ***Yogesh: #1 }}}
\newcommand{\aknote}[1]{  {\textcolor{blue}      { ***Alok: #1 }}}
\newcommand{\cwnote}[1]{  {\textcolor{brown}      { ***Charith: #1 }}}

}

\title{\emph{GoFFish}: A Sub-Graph Centric Framework for Large-Scale Graph Analytics\footnote{Under
    Review for a Conference, 2014}}

\author{%
Yogesh Simmhan$^\dagger$, Alok Kumbhare$^\star$, Charith Wickramaarachchi$^\star$,\\
Soonil Nagarkar$^\star$, Santosh Ravi$^\star$, Cauligi Raghavendra$^\star$, Viktor Prasanna$^\star$\\
\vspace{1.6mm}\\
{\fontsize{10}{10}\emph{$^{\dagger}$Indian Institute of Science, Bangalore 560012 India}}\\
{\fontsize{10}{10}\emph{$^\star$University of Southern California, Los Angeles CA 90089 USA}}
\vspace{1.2mm}\\
\fontsize{9}{9}\selectfont\ttfamily\upshape
simmhan@serc.iisc.in\\
\fontsize{9}{9}\selectfont\ttfamily\upshape
\{kumbhare, cwickram,  snagarka, sathyavi, raghu, prasanna\}@usc.edu
}

\begin{document}
\maketitle
\begin{abstract} 

Large scale graph processing is a major research area for Big Data exploration. Vertex centric
programming models like Pregel are gaining traction due to their simple abstraction that allows for
scalable execution on distributed systems naturally.  However, there are limitations to this
approach which cause vertex centric algorithms to under-perform due to poor compute to
communication overhead ratio and slow convergence of iterative superstep.  In this paper we introduce \emph{GoFFish} a
scalable sub-graph centric framework co-designed with a distributed persistent graph storage
for large scale graph analytics on commodity clusters. We introduce a \emph{sub-graph centric programming
abstraction} that combines the scalability of a vertex centric approach with the flexibility of
shared memory sub-graph computation. We map Connected Components, SSSP and PageRank algorithms to
this model to illustrate its flexibility. Further, we empirically analyze GoFFish using several real
world graphs and demonstrate its significant performance improvement, \emph{orders of magnitude in
some cases}, compared to Apache Giraph, the leading open source vertex centric implementation.

\end{abstract}

\section{Introduction}
\label{sec:intro}

The well-recognized challenge of ``Big Data'' manifests itself in many forms, ranging from data
collection and storage, to their analysis and visualization. Massive datasets from large scientific
instruments (e.g. Large Hadron Collider
, Large-Synoptic Sky Survey
) and
enterprise data warehouses (e.g. web logs, financial transactions) offered the initial grand
challenges for data-intensive computing that led to transformative advances such as
distributed and NoSQL databases, the MapReduce programming model and datacenters based on
commodity hardware. The unprecedented \emph{volume of data} was the dimension of ``bigness'' that was
addressed here.

With the proliferation of ubiquitous physical devices (e.g. urban monitoring, smart power
meters~\cite{lee2008cyber}) and virtual agents (e.g. Twitter feeds, Foursquare
check-ins
) that sense, monitor and track human and environmental activity,
the volume of intrinsically interconnected datasets is increasing. Indeed, a defining characteristic
of such datasets endemic to both the Internet of Things~\cite{atzori2010internet} and Social Networks is the
relationships that exist within them. Such graph datasets offer unique challenges to scalable Big
Data management and analysis even as they are becoming pervasive.



There has been significant work on parallel algorithms and frameworks for large graph applications
on HPC clusters~\cite{boost-graph}\footnote{http://www.graph500.org}, massively multithreaded shared memory
architecture~\cite{madduri2009}, and GPGPUs~\cite{harish2007accelerating}. Our focus in this paper, however, is
on leveraging 
\emph{commodity hardware} for scaling graph analytics, as such distributed
infrastructure, including cloud infrastructure, have democratized access to computing resources. This is evident from the
proliferation of Big Data programming models and frameworks, such as Hadoop/MapReduce~\cite{dean2008mapreduce}, for
composing and executing applications.  Besides resource access, another reason for
MapReduce's success~\cite{olston2008pig} is the simple programming model, which is constantly
evolving~\cite{ekanayake2010twister} and lowering the barrier to build scalable
applications. However, its tuple-based approach is ill-suited for many graph applications~\cite{lin2010mlg}

Recent platforms for graph analytics range from graph databases to \emph{vertex-centric 
programming abstractions}~\cite{malewicz2010pregel, low2012distributed}. In particular, Google's Pregel~\cite{malewicz2010pregel}
marries the ease of specifying a uniform application logic for each vertex with a \emph{Bulk Synchronous
Parallel (BSP)} execution model~\cite{valiant1990bmp}. Vertex executions that take place independently in a
distributed environment are interleaved with synchronized message exchanges across them to form
\emph{supersteps} that execute iteratively
Apache Giraph~\cite{avery2011giraph} is a popular open-source implementation of
Pregel~\cite{salihoglu2013gps} that has been scaled to trillions of edges
~\footnote{https://www.facebook.com/notes/facebook-engineering/scaling-apache-giraph-to-a-trillion-edges/10151617006153920}.

However, there are a few short-comings to this approach. (1) Operating on each vertex independently,
without a notion of shared memory, requires costly message exchanges at superstep boundaries. 
(2) Porting 
shared memory graph algorithms to a vertex centric one is also not simple, as
it limits algorithmic flexibility. A trivial mapping can have punitive performance. (3) The programming abstraction is decoupled from the
data layout on disk. Using na\"{\i}ve vertex distribution across machines can cause I/O penalties 
at initialization and at runtime. Recent research suggest some benefits from intelligent
partitioning~\cite{salihoglu2013gps}. 

In this paper, we propose a sub-graph centric programming abstraction, \emph{Gopher}, for performing
distributed graph analytics. This abstraction balances the flexibility of reusing well-known
shared-memory graph algorithms with the simplicity of a vertex centric iterative programming model
that elegantly scales on distributed environments. We couple
this abstraction with an efficient distributed storage, \emph{Graph oriented File System
(GoFS)}. GoFS partitions graphs across hosts while coalescing connected components within partitions
to optimize sub-graph centric access patterns. Gopher and GoFS are co-designed as part of the
\emph{GoFFish} framework to ensure that the data layout is intelligently leveraged during execution time.

We make the following specific contributions in this paper:
\begin{enumerate}
\item We propose a sub-graph centric programming model for composing graph analytics
  (\S~\ref{sec:gopher}), and map common graph applications to this model (\S~\ref{sec:algos}).
\item We present the GoFFish architecture which includes \emph{GoFS} for distributed
  graph storage and \emph{Gopher} for executing sub-graph centric applications (\S~\ref{sec:arch}).
\item We experimentally evaluate the proposed model for scaling common graph algorithms on real
  world graphs, and compare it against Giraph (\S~\ref{sec:eval}). 
\end{enumerate}



\section{Related Work}
\label{sec:related}



The popularity of MapReduce~\cite{dean2008mapreduce} for large scale data analysis has extended to
graph data as well~\cite{kang2008hadi}, with research techniques to scale it to peta-bytes of graph
data for some algorithms~\cite{kang2009pegasus, papadimitriou2008disco}. However, the
tuple-based approach of MapReduce makes it unnatural to develop graph algorithms, often requiring multiple
MapReduce stages~\cite{cohen2009cise}, additional programming constructs~\cite{chen2010sigmod} or
specialized platform tuning~\cite{lin2010mlg}. While the platform may scale, it is not the most
efficient or intuitive for many graph algorithms, often reading and writing the entire graph to disk
several times.

Message passing has been an effective model for developing scalable graph
algorithms~\cite{lumsdaine2007challenges}. In particular, the Bulk Synchronous Parallel (BSP) model
proposed by Valiant~\cite{valiant1990bmp} has been adopted by Google's
Pregel~\cite{malewicz2010pregel} for composing vertex centric graph algorithms. GraphLab takes a
similar vertex centric approach~\cite{low2012distributed}. Here, the computation logic is developed from the perspective of a single
vertex. The vertices operate independently (in parallel) and are initially aware only of their neighboring vertices. After the computation step, the vertices perform bulk message passing with their neighboring vertices (or other vertices
they discover).  Computation is done iteratively through a series of barriered supersteps that
interleave computation with bulk message passing~\cite{gerbessiotis1994direct}. This approach
naturally fits a distributed execution model, eliminates deadlock and race concerns of
asynchronous models, and simplifies the programming. 

Despite its growing popularity, the vertex centric model has its deficiencies.  Shared memory algorithms 
cannot be trivially mapped to such a vertex centric model and novel algorithms for standard graph
operations need to be developed.  For many graph algorithms the work performed on each vertex is negligible,
and the overhead of massive parallelization per vertex and the synchronization overhead per
superstep can out-weigh the benefits. 
Our sub-graph centric model of computing offers the additional flexibility of using shared
memory algorithms to deliver significant performance benefits.


Apache Giraph~\cite{avery2011giraph} is an open source implementation of Google's Pregel that is being adopted
by Facebook, among others. Alternatives such as Hama~\cite{seo2010hama},
Pregel.NET~\cite{redekopp2011performance} and GPS~\cite{salihoglu2013gps} also exist.  Giraph executes the
vertex centric program using Hadoop Map-only tasks. It also
offers additional programmatic extensions and engineering optimizations such as master-compute model, message
aggregation, and memory compression. Many of these focus on reducing the number of
messages passed since the messaging overhead is dictated by the
number of edges in the graph. For graphs with power-law edge distribution, highly connected vertices
can overwhelm the network and memory with messages. 

GPS~\cite{salihoglu2013gps} studies the effect of static partitioning on the Pregel model and
introduces runtime optimizations by performing dynamic partition balancing and replicating highly
connected vertices to neighbors. Our own work on Pregel.NET~\cite{redekopp2011performance} used a
swathe-based incremental scheduling to amortize the messaging overhead across staggered supersteps
for all-pairs algorithms like Betweenness Centrality and All Pairs Shortest Path.
While these optimizations by Giraph, GPS and Pregel.NET attempt to reduce inefficiencies through
engineering approaches, we posit in this paper that the vertex centric model in itself needs to be
enhanced to a sub-graph centric model to address critical performance deficiencies.

Besides Pregel, there are other distributed graph processing systems such as Trinity
\cite{shao2013trinity} that offer a shared memory abstraction within a distributed memory
infrastructure. Algorithms can use both message passing and a global distributed address space
called memory cloud. However, it assumes large memory machines with high speed interconnects. We focus on commodity cluster and do not make any assumptions about the underlying infrastructure.
Others such as Kineograph~\cite{cheng2012kineograph} 
focus
on realtime analysis of streaming graphs. 
Our focus here is on static graphs that are analyzed
offline on distributed commodity systems. Separately, there is ongoing work to support analytics
over timeseries of graphs using GoFFish.


There is a large body of work on parallel graph processing~\cite{lumsdaine2007challenges} for high
performance systems such as HPC clusters~\cite{boost-graph}, massively multithreaded architectures like the
Cray XMT~\cite{madduri2009,Ediger2013ipdpsphd} and GPGPUs~\cite{harish2007accelerating}. Our focus in this paper is on commodity
clusters and infrastructure Clouds that are interconnected by high-latency Ethernet, spinning disks
and no shared memory. We evaluate the scalability of our framework on such accessible commodity
hardware, and do not compete with high end infrastructure or runtimes optimized for
them~\cite{murphy2010introducing} However, parallel graph algorithms developed for these other platforms may
still be relevant. Indeed, the BSP model we leverage was developed by Valint for parallel computing
in 1980.


\section{Sub-graph Centric Programming Abstraction}
\label{sec:gopher}

\subsection{Vertex Centric Programming Model \& Gaps}

Vertex centric programming models like Pregel offer a simple abstraction for composing graph
algorithms for distributed systems. Vertices in the graph are partitioned across
machines. Users implement a \texttt{Compute} method that gives them access to a single vertex and
its value(s), the outgoing edge list for the vertex, and the ability to send custom messages to
vertices that these edges are incident upon. The execution model is iterative, based on a Bulk
Synchronous Parallel (BSP) paradigm~\cite{valiant1990bmp}. The \texttt{Compute} method is executed
independently for each vertex and messages it generates to neighboring vertices are available to
them only after \texttt{Compute} completes for all vertices. This forms one \emph{superstep}. A
barrier synchronization at the end of a superstep ensures that all generated messages are delivered
to destination vertices before the next superstep's \texttt{Compute} method is invoked for each
vertex with the input messages. The vertices can \texttt{VoteToHalt} in their \texttt{Compute}
method; any vertex that has voted to halt is not invoked in the next superstep unless it has input
messages. The application terminates when all vertices have voted to halt and there are no new input
messages available in a superstep.

Pseudocode to find the maximum value among all vertices is shown in
Algorithm~\ref{alg:pregel-maxvalue}. Each vertex sends its value to neighboring vertices in the
first superstep (\textsc{SendToAllNeighbors}). In later supersteps, it updates its value to the highest value among all incoming
messages. If changed, it sends its value to its neighbors; otherwise, it votes to halt. The
application terminates when all vertices have reached a steady state value, which equals the largest
vertex's value.

\begin{algorithm}
\small
\caption{Max Vertex using Vertex Centric Model}\label{alg:pregel-maxvalue}
\begin{algorithmic}[1]
\Procedure{Compute}{Vertex myVertex, Iterator$\langle$Message$\rangle$ M}
\State hasChanged = (superstep == 1) ? \texttt{true} : \texttt{false}
\While{M.hasNext} \Comment{\emph{Update to max message value}}
\State Message m  $\gets$ M.next
\If{m.value $>$ myVertex.value}
\State myVertex.value $\gets$ m.value
\State hasChanged = \texttt{true}
\EndIf
\EndWhile
\If{hasChanged} \Comment{\emph{Send message to neighbors}}
\State \Call{SendToAllNeighbors}{myVertex.value}
\Else
\State \Call{VoteToHalt}{ }
\EndIf
\EndProcedure
\end{algorithmic}
\end{algorithm}

Despite the simplicity of programming, there are two key scalability bottlenecks in this vertex centric
approach: (1) the number of messages exchanged between the vertices, and (2) the number of
synchronized supersteps required for completion. Typically, vertices are hashed and distributed
across multiple machines (assuming one worker per machine). Message passing is done either
in-memory (for vertices on the same machine) or over the network, while barrier synchronization
across vertices on distributed machines is centrally coordinated. 
Vertices can often perform only light computations, making them communication bound.
Network messaging 
is costly given the commodity hardware of data centers, and not every algorithm can benefit
from \texttt{Combiners}~\cite{malewicz2010pregel} to reduce messages generated within a worker. The default mapping of
vertices to machines using (random) hashing exacerbates this though better \emph{a priori} partitioning shows only
limited improvements~\cite{salihoglu2013gps}. Even when passing messages in-memory to co-located vertices, the
number of intermediate messages between superstep can overwhelm the physical memory, requiring disk
buffering for algorithms such as Betweenness Centrality~\cite{redekopp2011performance}. At the same time, the
number of supersteps taken by a vertex centric algorithm can be large. Some
like PageRank may use a fixed number of supersteps (e.g. 30)~\cite{page1999pagerank} while others such as Max
Vertex are bound by the diameter of the graph. For e.g., the LiveJournal social network
with 
$\sim$4.8M vertices has edge degrees with powerlaw distribution and a diameter of 16 while the
California road network with 
$\sim$1.9M vertices has a diameter of 849 (Table~\ref{table:datasets}). The synchronization time is cumulative over supersteps and can thus be significant.

\subsection{Sub-graph Centric Programming}
\label{sec:subgraph-prog}

\begin{figure}
\centering
\includegraphics[width=0.9\columnwidth]{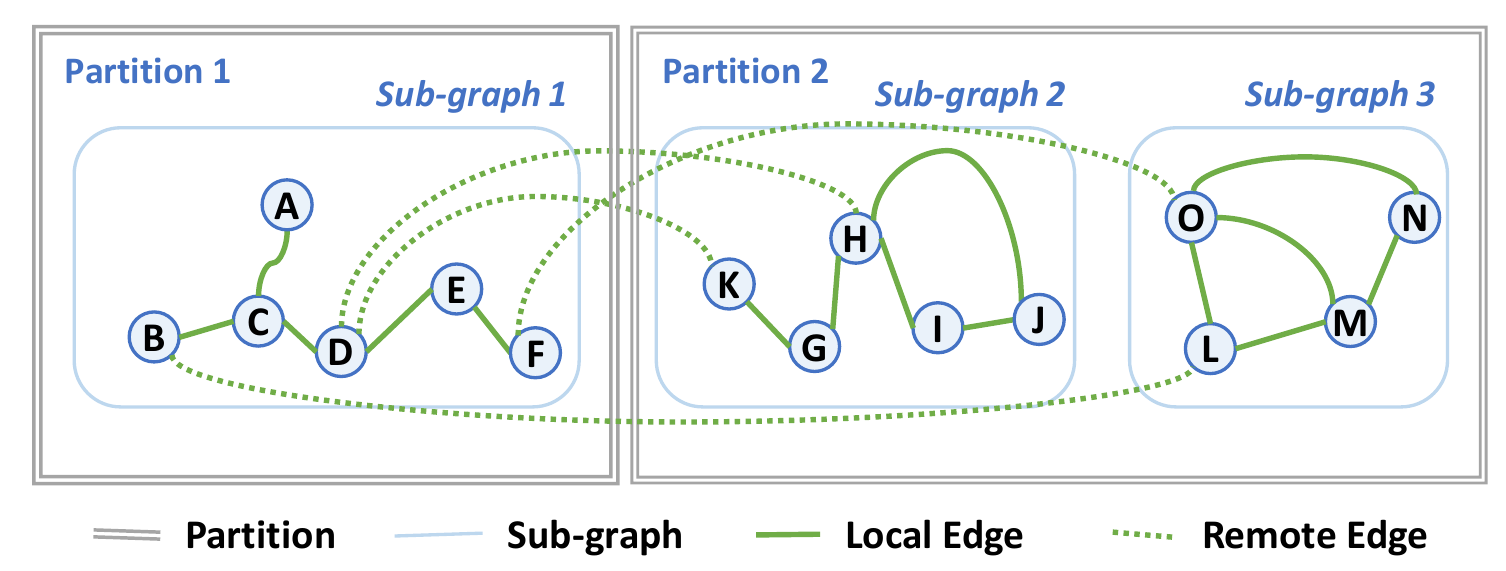}
\caption{Graph partitioned into two with \emph{sub-graphs} identified in each partition.}
\label{fig:subgraph}
\vspace{-0.2in}
\end{figure}

We propose a sub-graph centric programming abstraction that targets these deficiencies. As with
Pregel, we operate in a distributed environment where the graph is $k-way$ partitioned over its
vertices across $k$ machines. 
Further, we define a \textbf{sub-graph} as a connected component within
a partition of an undirected graph; they are weakly connected if the graph is directed (Fig.~\ref{fig:subgraph}).

Let $ P_{i} = \lbrace \mathbb{V}_{i},\mathbb{E}_{i}\rbrace $ be a graph partition $i$ where $V_{i}$
and $E_{i}$ are the set of vertices and edges in the partition. We define a \emph{sub-graph} S in
$P_{i}$ as $S = \lbrace V,E,R | v \in V \Rightarrow v \in
\mathbb{V}_{i}; e \in E \Rightarrow  e \in \mathbb{E}_{i}; r \in R \Rightarrow  r
\not\in \mathbb{V}_{i};  \forall u,v \in V \exists$ an undirected path between $u,v ;$
and $\forall r \in R $ $\exists v \in V$  s.t. $ e=\langle v,r \rangle \in E \rbrace$
where V is a set of local vertices, E is a set of edges and R is a set of remote vertices.
  
\begin{figure}
\includegraphics[width=\columnwidth]{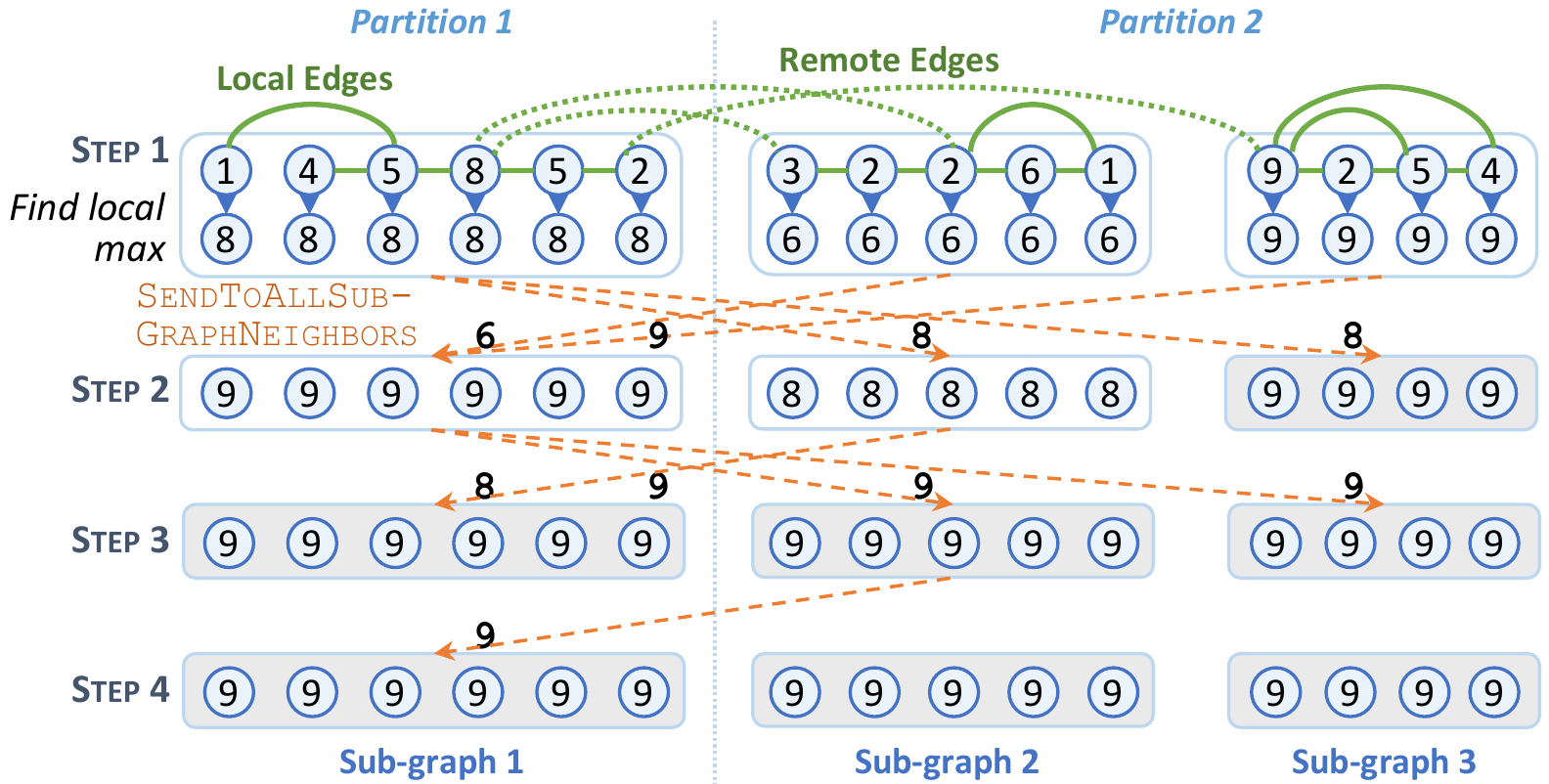}
\caption{
  Sub-graph centric maximum value computation for graph from Fig.~\ref{fig:subgraph}. Top row shows
  the graph at start with vertices having the initial value. Each row of rectangles is one superstep; each rectangle is an independent sub-graph
  computation. Dashed arrows show messages passed at superstep boundaries. 
  Vertices in a sub-graph
  have shared memory; those across sub-graphs do not. Vertex values are shown
  for the end of each superstep. Shaded sub-graphs have voted to halt.}
\label{fig:model}
\end{figure}

Two sub-graphs do not share the same vertex but can have remote edges that connect their vertices, as
long as the sub-graphs are on different partitions. If two sub-graphs on the same partition share an
edge, by definition they are merged into a single sub-graph. 
A single partition can have one or more sub-graphs
and the set of all sub-graphs forms the complete graph. Specific partitioning approaches are
discussed later, and each machine holds one partition. In other words, sub-graphs behave like ``meta
vertices'' with remote edges connecting them across partitions. For e.g., Fig.~\ref{fig:model} shows
an undirected graph with 15~vertices partitioned into two, with a total of three sub-graphs having
6, 5 and 4 vertices. Sub-graphs 1 and 2, and sub-graphs 1 and 3 share remote vertices.

As with a vertex centric model, each sub-graph is treated as an independent unit of computation
within a superstep. Users implement the following method signature:
\begin{center}\small 
\texttt{\textbf{Compute}(Subgraph, Iterator<Message>)}
\end{center} 

The \texttt{Compute} method can access the sub-graph topology and values of the
vertices and edges. The values are mutable though the topology is constant. This allows us to
fully traverse the sub-graph up to the boundary remote edges \emph{in-memory, within a single superstep} and
accumulate values of the sub-graph or update values for the local vertices and edges. There is no
shared memory across sub-graphs. Instead, they communicate by message passing, with messages being exchanged at synchronized
superstep boundaries, as with the BSP model. Several methods enable this messaging.

Algorithms often start by sending messages 
to neighboring sub-graphs. These are by definition on remote partitions.
\begin{center}\small 
\texttt{\textbf{SendToAllSubGraphNeighbors}(Message)}\\
\end{center}

\noindent As other sub-graphs are discovered, as part of the traversal and as
messages propagate, two other methods are useful:
\begin{center}\small 
\texttt{\textbf{SendToSubGraph}(SubGraphID, Message)}\\
\texttt{\textbf{SendToSubGraphVertex}(SubGraphID, VertexID, Message)}\\
\end{center}

\noindent We also allow a global broadcast to all sub-graphs, though this is costly and should be used sparingly.
\begin{center}\small 
\texttt{\textbf{SendToAllSubGraphs}(Message)}
\end{center}

As with a vertex
centric approach, the \texttt{Compute} method can invoke \texttt{\textbf{VoteToHalt}()} on the
sub-graph and the application terminates when all sub-graphs have halted and there are no new input
messages.

Algorithm~\ref{alg:gopher-maxvalue} presents the sub-graph centric version for finding the maximum
value among all vertices in a graph and Fig.~\ref{fig:model} shows its execution. Instead of a
vertex, the \texttt{Compute} method operates on a sub-graph \emph{mySG}. Lines~\ref{alg:line:step0start}--\ref{alg:line:step0end} are executed only
for the first superstep, where all sub-graph's value are initialized to largest value among its vertices. Subsequently, the algorithm is similar to the vertex centric version: we send the
sub-graph's value it its neighboring sub-graphs and update the sub-graph's value to the highest
value received, halting when there is no further change. At the end, each sub-graph has the value of
largest vertex.

Compared to the vertex centric algorithm, the sub-graph centric version reduces the number of
supersteps taken since the largest value discovered at any superstep propagates through
the entire sub-graph in the same superstep. For e.g., for the graph in Fig.~\ref{fig:model}, the
vertex centric approach takes 7 supersteps while we use 4 
. In addition to
reducing the superstep synchronization overhead, this also reduces the cumulative number of messages
exchanged over the network.  In the worst case, when a sub-graph is trivial (has just one vertex),
we degenerate to a vertex centric model.

\begin{algorithm}
\small
\caption{Max Vertex using Sub-Graph Centric Model}\label{alg:gopher-maxvalue}
\begin{algorithmic}[1]
\Procedure{Compute}{SubGraph mySG, Iterator$\langle$Message$\rangle$ M}
\algrenewcommand{\alglinenumber}[1]{\colorbox{light-gray}{\footnotesize\textbf{#1}:}}
\If{superstep = 1} \Comment{\emph{Find local max in subgraph}} \label{alg:line:step0start}
\State mySG.value $\gets -\infty$
\ForAll{Vertex myVertex \textbf{in} mySG.vertices}
\If{mySG.value $<$ myVertex.value}
\State mySG.value $\gets$ myVertex.value 
\EndIf
\EndFor
\EndIf \label{alg:line:step0end}
\algrenewcommand{\alglinenumber}[1]{\color{black}\footnotesize#1:}
\State hasChanged = (superstep == 1) ? \texttt{true} : \texttt{false} \label{alg:line:stepNstart}
\While{M.hasNext}
\State Message m  $\gets$ M.next
\If{m.value $>$ mySG.value}
\State mySG.value $\gets$ m.value
\State hasChanged = \texttt{true}
\EndIf
\EndWhile
\If{hasChanged}
\State \Call{SendToAllSubGraphNeighbors}{mySG.value}
\Else
\State \Call{VoteToHalt}{ }
\EndIf \label{alg:line:stepNend}
\EndProcedure
\end{algorithmic}
\end{algorithm}

\subsection{Benefits}
More generally, this simple and elegant extension from the vertex centric model offers several important benefits. 

\emph{1) Messages Exchanged.}
Access to the entire sub-graph enables the application to make a significant progress within each
superstep while reducing costly message exchanges that cause network transfer overhead, disk
buffering costs or memory pressure, between supersteps.  
Even if messages are generated for every
remote edge from a sub-graph, messages destined to the same sub-graph can be intelligently grouped
to mitigate network latency. While Pregel allows \texttt{Combiners} to be defined per worker, their
ability to operate only on the generated messages limits their utility. Access to the entire
sub-graph topology and values allows for more flexible and efficient algorithms that inherently reduce the number of generated messages. 

\emph{2) Number of Supersteps.}
Depending on the type of graph application, a sub-graph centric model can also reduce the number of
supersteps required to complete the application. This can reduce the time spent on superstep
synchronization. Further, the time taken by a superstep is based on its slowest worker and there can
be a wide distribution between the fastest and slowest workers~\cite{redekopp2011performance}. Reducing the number of supersteps
mitigates the cumulative impact of this skew. For traversal based algorithms, the number of
supersteps is a function of the longest shortest-path distance between any two vertices in the
graph 
i.e. the \emph{diameter} of the graph. Intuitively, the diameter corresponds to the maximal height
of the BFS tree rooted at any vertex in the graph. In a sub-graph centric model, this
becomes the diameter of the meta-graph where the sub-graphs form meta-vertices. In the best case (a linear
chain), the number of supersteps can reduce by a factor proportional to the number vertices in the sub-graph, while at
the other extreme (trivial sub-graph), it is no worse than the number of supersteps taken for a
vertex-centric model. We should note that this improvement may be nominal for 
small-world networks with small diameters, and also for non-traversal based 
algorithms, like PageRank, as we discuss later. 

\emph{3) Reuse of Single-machine Algorithms.}
Lastly, the programming overhead introduced by a sub-graph centric approach relative to a
vertex-centric approach is incremental. Often, it involves using a shared-memory graph
algorithm on the sub-graph (e.g. Algorithm~\ref{alg:gopher-maxvalue},
lines~\ref{alg:line:step0start}--\ref{alg:line:step0end} with shaded line numbers) and then
switching to an approach similar to a vertex-centric one but treating sub-graphs as vertices
(lines~\ref{alg:line:stepNstart}--\ref{alg:line:stepNend})). Given that efficient graph algorithms
for single machines are known, even leveraging many cores or accelerators, the potential
impact on programming ease is offset by the increase in programming flexibility and performance.


\section{Architecture}
\label{sec:arch}


\emph{GoFFish} is a scalable software framework for storing graphs, and composing and executing
graph analytics in a Cloud and commodity cluster environment~\footnote{\url{https://github.com/usc-cloud/goffish}}. A \emph{Graph oriented File System (GoFS)}
and \emph{Gopher execution engine} are co-designed \emph{ab initio} to ensure efficient
distributed storage for sub-graph centric data access patterns during loosely coupled execution.
The design choices target commodity or virtualized hardware with Ethernet and spinning disks rather
than HPC environments with low latency networking and disks. GoFFish is implemented in Java.

\begin{figure}
\includegraphics[width=\columnwidth]{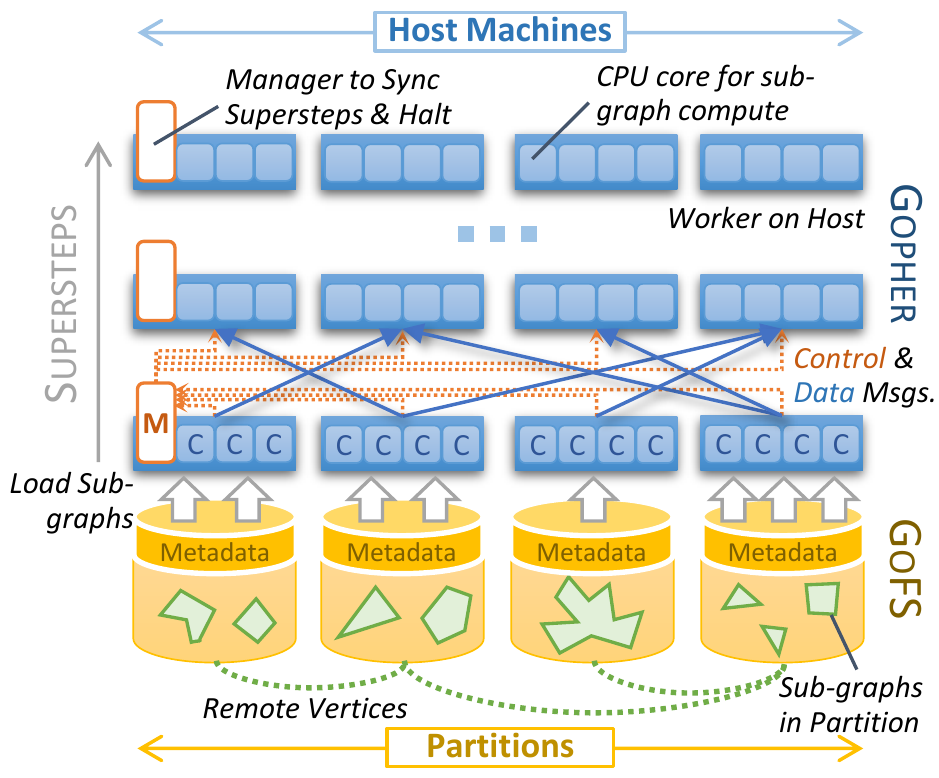}
\caption{Bulk Synchronous Execution of Sub-graph centric computation across supersteps.}
\label{fig:arch}
\end{figure}

\subsection{GoFS Distributed Graph Store}
GoFS is a distributed store for partitioning, storing and accessing graph datasets across hosts in a
cluster.  Graphs can have both a \emph{topology} and \emph{attributes} associated with each vertex
and edge. The former is an adjacency list of uniquely labeled vertices and (directed or undirected)
edges connecting them. Attributes are a list of name-value pairs with a schema provided for typing.
Input graphs are partitioned across hosts, one partition per machine, using the METIS
tool~\cite{karypis1995metis} to balance vertices per partition and minimize edge cuts (Fig.~\ref{fig:arch}).

GoFS uses a sub-graph oriented model for mapping the partition's content to \emph{slice files},
which form units of storage on the local file system. We identify all sub-graphs in the partition
-- components that are (weakly) connected through local edges, and a partition with $n$ vertices can
have between 1 to $n$ sub-graphs. Each sub-graph maps to one \emph{topology slice} that contains
local vertices, local edges and remote edges, with references to partitions holding the destination
remote vertex, and several \emph{attribute slices} that hold their names and values. We use
\emph{Kryo} 
to efficiently convert slices objects into a compact binary form on
file with smaller disk access costs.

GoFS is a \emph{write once-read many} scalable graph data store rather than a database with rich
update and query support. The GoFS Java API allows clients to access a graph's metadata, attribute
schema and sub-graphs present in the local partition. Specific sub-graphs and select attributes can
be loaded as a Java object for in-memory traversals of local vertices and edges. Remote edges in a
sub-graph resolve to a remote partition, sub-graph and vertex ID that are then accessed locally on
that machine. 


\subsection{Gopher Sub-graph Centric Framework}
Gopher is a programming framework that offers our proposed sub-graph centric abstractions
(Sec.~\ref{sec:subgraph-prog}), and executes them using the \emph{Floe} \cite{floe} dataflow engine on a Cloud
or cluster in conjunction with GoFS. 
It inherits the natural scalability of the vertex centric model but extends it further. 
Users implement their algorithm in Java within a
\texttt{Compute} method where they get access to a sub-graph and data messages from the previous
superstep. They can use the \texttt{Send*} methods to send messages to sub-graphs in the next
superstep and call the \text{VoteToHalt} method. The same \texttt{Compute} method logic is executed
on every sub-graph in the graph, on all partitions, for each superstep.

The Gopher framework has a \emph{compute worker} service running on each machines in the cluster and
a additional \emph{manager} service on one machine. Gopher deploys the user's compute logic to the
workers to operate on a particular graph. The workers initially load all local sub-graphs for that
graph into memory using GoFS. For every superstep, it uses a thread pool optimized for multi-core
CPUs to invoke the \texttt{Compute} on each sub-graph. \texttt{Send*} data messages emitted by the
\texttt{Compute} are resolved by GoFS to the remote partition and host. The worker aggregates
messages destined for the same host and sends them over TCP sockets to the remote
worker asynchronously as the compute continues to run.

When the \texttt{Compute} for all sub-graphs in a partition complete in a superstep, the worker
flushes all pending messages to the remote workers and then sends a \emph{synchronization} control
message to the manager. Once the manager receives sync messages from all workers, it broadcasts a
\emph{resume} control message to the workers to start their next superstep. The workers then call the
\texttt{Compute} method again on their sub-graphs and also pass them a list of data messages
received from the previous superstep. The method is stateful for each sub-graph; so local variables
are retained across superstep. \texttt{Compute} is called for a sub-graph only if (1) it has not
voted to halt in the previous superstep and (2) it has received no input messages. When a worker
does not have to call \texttt{Compute} for any of its sub-graphs in a superstep, it sends a
\emph{ready to halt} control message to the manager. When all workers are ready to halt at a
superstep, the manager send a terminate \emph{control} message for the workers to stop.

\subsection{Storage-Compute Co-design}

Designing data layout and execution models together helps data-intensive applications, as seen with
Hadoop and HDFS. Unlike their tuple-based model, GoFFish emphasizes sub-graphs as a logical unit of
storage and compute. Hence our data distribution is based on graph partitioning followed by
sub-graph discovery. While there is a higher overhead in loading data into GoFS, this layout
minimizes runtime network data movement in Gopher. The distributed layout also maximizes cumulative
disk read bandwidth across machines when loading sub-graphs into Gopher. We also balance the disk
latency (\# of unique files read) against sequential bytes read: having separate slice files for a
sub-graph's topology and attributes allows only those slices to be read from disk, but they can be
read in their entirety. For e.g. a graph with 10 attributes on its vertices and edges but using only
the edge weight for a Gopher algorithm needs to only load that slice 

Partitioning can impact the performance of several sub-graph centric algorithms we have analyzed
and will show empirically. Since we use out-of-the-box tools like METIS for partitioning, these help
balance vertices and minimize edge cuts. However, ideally, we should be balancing the number of
sub-graphs across partitions and have uniform sizes, in addition to reducing edge cuts. This can
help workers complete a superstep in the same time. 
Also, if the number of sub-graphs in a partition
is a multiple of the number of cores in a machine, we can optimally leverage the parallelism. 
While some of these partitioning schemes are for future work, the design sets clear goals for this
optimization.  





\section{Sub-Graph Centric Graph Algorithms}
\label{sec:algos}

We present several sub-graph centric algorithms for common graph analytics, both to illustrate the
utility and discuss the benefits of our proposed abstraction. We also highlight the algorithmic
optimizations that are possible using a sub-graph centric model, relative to a vertex centric one.



\subsection{Connected Components}
Connected components identify maximally connected sub-graphs within an undirected graph such that
there is path from every vertex to every other vertex in the sub-graph. The sub-graph
centric algorithm (and the vertex centric one) for connected components is similar to the Maximum Value
algorithm (Algorithm~\ref{alg:gopher-maxvalue}). It in itself is
based on the HCC algorithm~\cite{kang2009pegasus}. Briefly,
assuming each vertex has a unique identifier, we propagate the ID of the vertex with the largest ID value within a sub-graph to its connected neighbors.
In effect, we perform a breadth first traversal rooted at the sub-graph with the largest vertex ID, 
with each superstep traversing one more level till the farthest connected sub-graph is reached. We will eventually
have vertices labeled with the component ID (i.e. largest vertex ID) they belong to. The number of unique vertex values gives
us the number of components.

The computational complexity of this
algorithm is 
$ O((d+1) \times v/p)$, where $d$ is the diameter of the graph (specifically, of the largest connected component) constructed by treating
each sub-graph as a meta vertex, $v$ is the number of vertices in the graph and $p$ is the degree of
parallelism. $p$ equals the number of
machines (or partitions) in the cluster, assuming the number of vertices per machine is balanced. In fact, if
there are at least as many uniformly sized sub-graphs per partition as the number
of cores, the value of $p$ approaches the total number of
cores in the cluster. 
The key algorithmic optimization in this case comes from reducing the number of
supersteps $(d+1)$ relative to the vertex centric model.



\subsection{Single Source Shortest Path (SSSP)}
Single Source Shortest Path (SSSP) determines the shortest distance from a source vertex to every
other vertex in the graph. Intuitively, the sub-graph centric algorithm finds the shortest distances
from the source vertex to all internal vertices (i.e. not having a remote edge) in the sub-graph
holding the source in one superstep using \textsc{Dijkstras} (Algorithm~\ref{alg:gopher-sssp}). It then sends the updated distances
from the vertices having a remote edge to their neighboring sub-graphs. These
sub-graphs propagate the changes internally, in one superstep, and to their neighbors, across
supersteps, till the distance values quiesce.


The time complexity of this algorithm is a function of the time taken to perform \textsc{Dijkstras}
on the sub-graphs in a superstep and the number of supersteps. Using a priority queue in
\textsc{Dijkstras}, we take $O(log(v))$ for vertex lookups, and this is performed for each of $e$ edges in a
sub-graph with $v$ vertices. This gives a compute time of $O( (e \cdot log(v)))$ per
superstep, where $e$ and $v$ are typically dominated by
the largest active sub-graph in a superstep. The number of supersteps is a function of the 
graph diameter $d$ measured through
sub-graphs, and this takes $O(d)$ supersteps.  
For a partitions having a large number of small sub-graphs, we
need to multiple this by a factor $s/c$ where $s$ is the number of sub-graphs in a machine and $c$
the level of concurrency (i.e. number of cores) on that machine. As compared to a vertex centric
approach, the time complexity per superstep is larger since we run \textsc{Dijkstras} (rather than
just update distances for a single vertex), but we may significantly reduce the number of supersteps
taken for the algorithm to arrive at the shortest vertex distances.
\begin{algorithm}
\small
\caption{Sub-Graph Centric Single Source Shortest Path}\label{alg:gopher-sssp}
\begin{algorithmic}[1]
\Procedure{Compute}{SubGraph mySG, Iterator$\langle$Message$\rangle$ M}
\State{openset $\gets \varnothing $} \Comment{\emph{Vertices with improved distances}}
\If{superstep = 1} \Comment{\emph{Initialize distances}}
\ForAll{Vertex v \textbf{in} mySG.vertices}
\If {v = SOURCE}    
\State {v.value $\gets$ 0} \Comment{\emph{Set distance to source as 0}}
\State {openset.add(v)} \Comment{\emph{Distance has improved}}
\Else
\State {v.value $\gets -\infty$} \Comment{\emph{Not source vertex}}
\EndIf
\EndFor
\EndIf

\ForAll{Message m \textbf{in} M} \Comment{\emph{Process input messages}}
\If{mySG.vertices[m.vertex].value $>$ m.value} 
\State{mySG.vertices[m.vertex].value $\gets$ m.value}
\State{openset.add(m.vertex)} \Comment{\emph{Distance improved}}
\EndIf
\EndFor

\State{remoteSet $\gets$ \Call{Dijkstras}{mySG, openset}}

\State \Comment{\emph{Send new distances to remote sub-graphs/vertices}}
\ForAll{$\langle$remoteSG,vertex,value$\rangle$ \textbf{in}
  remoteSet}
\State{\Call{SendToSubGraphVertex}{remoteSG, vertex, value}}
\EndFor

\State{\Call{VoteToHalt}{ }}

\EndProcedure

\Statex

\Procedure{Dijkstras}{Set$\langle$Vertex$\rangle$ openset}
\State remoteOpenset $\gets \varnothing$
\While{ openset $\neq \varnothing$}
\State  Vertex v $\gets$ \Call{GetShortestVertex}{openset}
\ForAll{Vertex v2 \textbf{in} v.neighbors} 
\State \Comment{\emph{Update neighbors, notify if remote.}}
\If{v2.isRemote()}
\State  remoteOpenset.add(v2.subgraph, v2, v.value + 1)
\ElsIf{v2.value $>$ v.value + 1}
\State        v2.value $\gets$ v.value + 1 
\State        openset.add(v2) \Comment{\emph{Distance has improved}}
\EndIf
\EndFor 
\State  openset.remove(v) \Comment{\emph{Done with this local vertex}}
\EndWhile
\State \Return remoteOpenset
\EndProcedure

\end{algorithmic}
\end{algorithm}

\subsection{PageRank and BlockRank}
PageRank calculates the rank of a web page based on the probability with which an 
idealized random web surfer 
will end
on that page~\cite{page1999pagerank}. For each superstep in a vertex centric model~\cite{malewicz2010pregel}, a vertex adds all input
message values into $sum$, computes 
$0.15 / v + 0.85 \times sum$ as its new value, and sends
$value/g$ to its $g$ neighbors. The $value$ is $1/v$ initially, for $v$ vertices in the graph. 
An equivalent sub-graph centric approach does not confer algorithmic benefits; it takes the same
$\sim$30 supersteps to converge and each vertex operates independently in lock step, with an $O(30
\cdot \frac{v}{p \cdot c} \cdot g )$, for $g$ average edge degree.  

The
\emph{BlockRank} 
algorithm, however, uses the property that some websites
are highly inter-connected (i.e. like sub-graphs) to set better initial vertex values that hasten
convergence. It calculates the PageRank for vertices by treating blocks (sub-graphs) independently
\emph{(1 superstep)}; ranks each block based on its relative importance 
\emph{(about 1 superstep)}; and normalizes the
vertex's PageRank with the BlockRank to use as an initial seed before running \emph{classic PageRank} \emph{($n$
supersteps)}. The first superstep is costlier as the sub-graphs estimate PageRanks for their
vertices, but converges in fewer (local) iterations~\cite{kamvar2003exploiting}. Similarly, the last $n$ supersteps
too converge more quickly than 
$\sim$30, thus using fewer supersteps than classic
PageRank.

%





\subsection{Discussion}


In summary, the relative algorithmic benefits of using a sub-graph centric abstraction can be
characterized based on the class of graph algorithm and graph.  For algorithms that perform
full graph traversals, like SSSP, BFS and Betweenness Centrality, we reduce the number of supersteps to a
function of the diameter of the graph based on sub-graphs rather than vertices. This can offer
significant reduction. However, for powerlaw graphs that start with a small vertex based diameter,
these benefits are muted. For algorithms that use fixed numbers of iterations, like classic
PageRank, there may not be any reduction unless we use alternatives like BlockRank. 

The time complexity per superstep can be larger since we often run the single-machine graph
algorithm on each sub-graph. The number of vertices and edges in large sub-graph will impact this. 
If there are many small sub-graphs in a partition, 
the number of sub-graphs becomes the limiting
factor as we approach a vertex centric behavior.  For graphs with high edge density, algorithms
that are a linear (or worse) function of the number of edges can take longer supersteps. Also, for
algorithms like SSSP, such graphs offer more paths through remote vertices thus taking more supersteps
to converge.



\section{Performance Analysis}
\label{sec:eval}
We evaluate the GoFFish framework against Apache Giraph
, a popular open source implementation
of Google's Pregel
vertex centric programming model that uses HDFS storage. 
We use the latest development version of Giraph, at
the time of writing, which includes performance enhancements like lazy
de-serialization, staggered aggregators, and multi-threaded data reads contributed by
Facebook
. 
We analyze the relative performance of sub-graph centric Gopher and vertex centric Giraph for representative graph
algorithms: \emph{Connected Components}, \emph{Single Source Shortest Path (SSSP)} and (classic)
\emph{PageRank}, as discussed in Sec.~\ref{sec:algos}.

\subsection{Experimental Setup and Datasets}
We run these experiments on a cluster of 12~nodes, each with an 8-core Intel Xeon CPU, 16~GB RAM and
1~TB SATA HDD, and connected by Gigabit Ethernet. Both Giraph and GoFFish are deployed on all nodes.
Giraph is configured with the default two workers per node and 8~GB RAM per worker. GoFFish runs a
single worker per node with 16~GB RAM, with one node also hosting the manager. Both run on Oracle
Java~7 JRE for 64~bit Ubuntu Linux. 

\begin{table}
\centering
\begin{tabular*}{\columnwidth}{p{0.25\columnwidth}||r|r|r|r}
\hline
\textbf{Dataset}&\emph{Vertices}&\emph{Edges}&\emph{Diameter}&\emph{WCC}\\
\hline
\hline
\textbf{RN}: California Road Network
& {\raggedleft 1,965,206} & {\raggedleft 2,766,607} & 849 & 2,638 \\
\textbf{TR}: Internet from Traces & 19,442,778 & 22,782,842 & 25 & 1 \\
\textbf{LJ}: LiveJournal Social Network
& 4,847,571 & 68,475,391 & 10 & 1,877 \\
\hline
\end{tabular*}

\caption{Characteristics of graph datasets used in evaluation}
\label{table:datasets}
\vspace{-0.2in}
\end{table}


We choose a range of real world graphs (Table~\ref{table:datasets}): California road network
(\textbf{RN}), Internet route path network constructed from a CDN traceroute dataset
(\textbf{TR})), and a snapshot of the LiveJournal social network (\textbf{LJ}). These
graphs are diverse. RN is a relatively small, sparse network with an even distribution of
small edge degrees and a large diameter. LJ, on the other hand, is dense graph with powerlaw edge
degrees and a small diameter. TR has a powerlaw edge degree too, with a small number of highly
connected vertices (ISPs and a vertex used to indicate trace timeouts). 

Unless otherwise stated, we report average values over three runs each for each experiment. The
entire cluster is dedicated to either GoFFish or Giraph during a run.

\subsection{Summary Results}
\begin{figure*}[t]
\centering
\subfigure[Total time (log scale) for GoFFish and Giraph for various graph algorithms and datasets]{
  \includegraphics[width=0.75\textwidth]{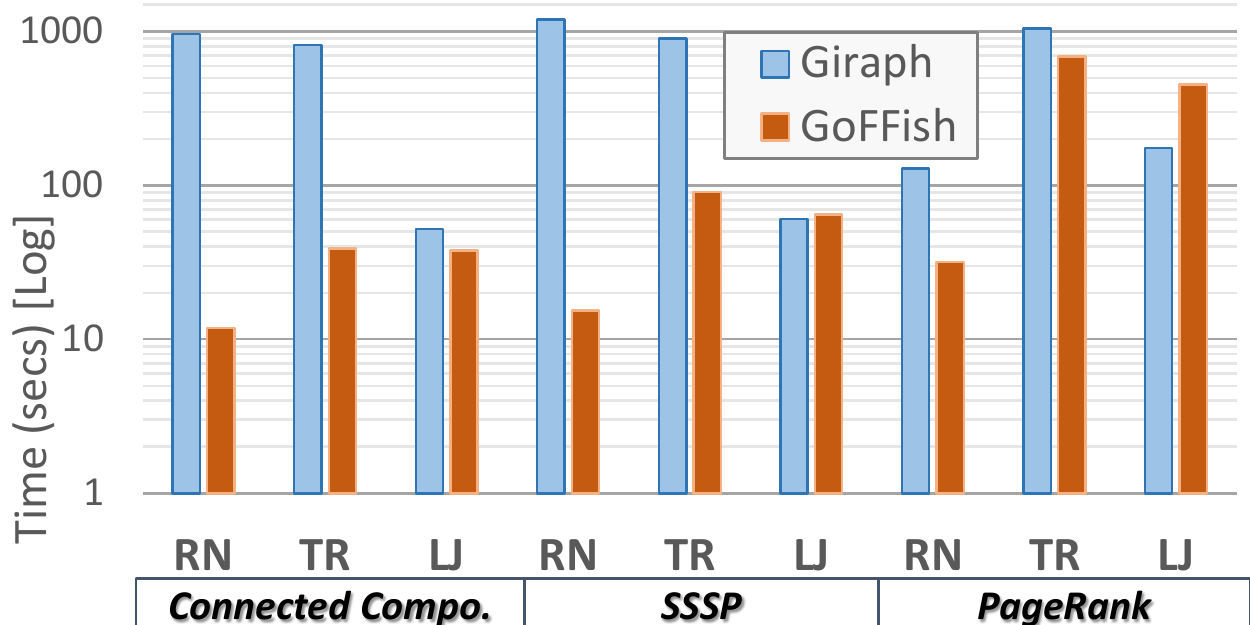}
  \label{fig:total}
}\\
\subfigure[Graph loading time (log scale) from disk to memory. \emph{Edge Imp.} is GoFFish with
load improvements.]{
	\includegraphics[width=0.44\textwidth]{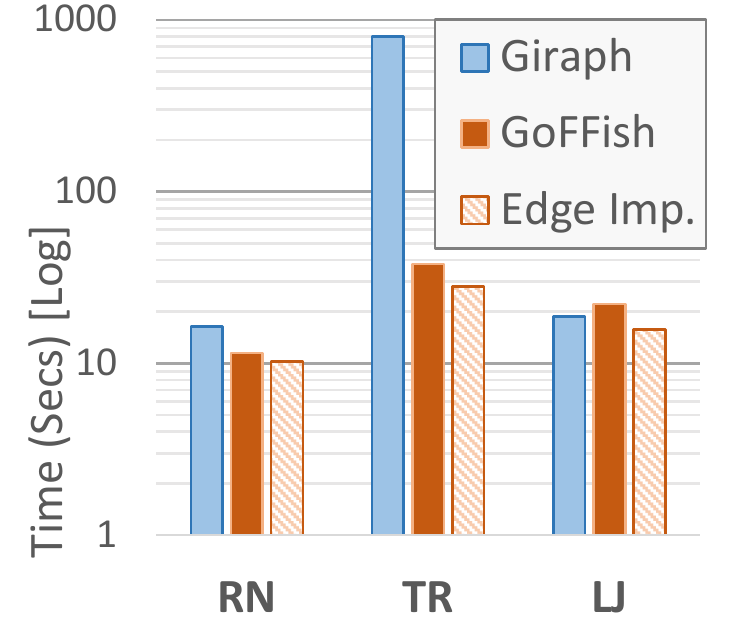}
	\label{fig:dataload}
}\\
\subfigure[Number of supersteps (log scale) taken by Gopher and Giraph for various graph algorithms and datasets]{
	\includegraphics[width=0.75\textwidth]{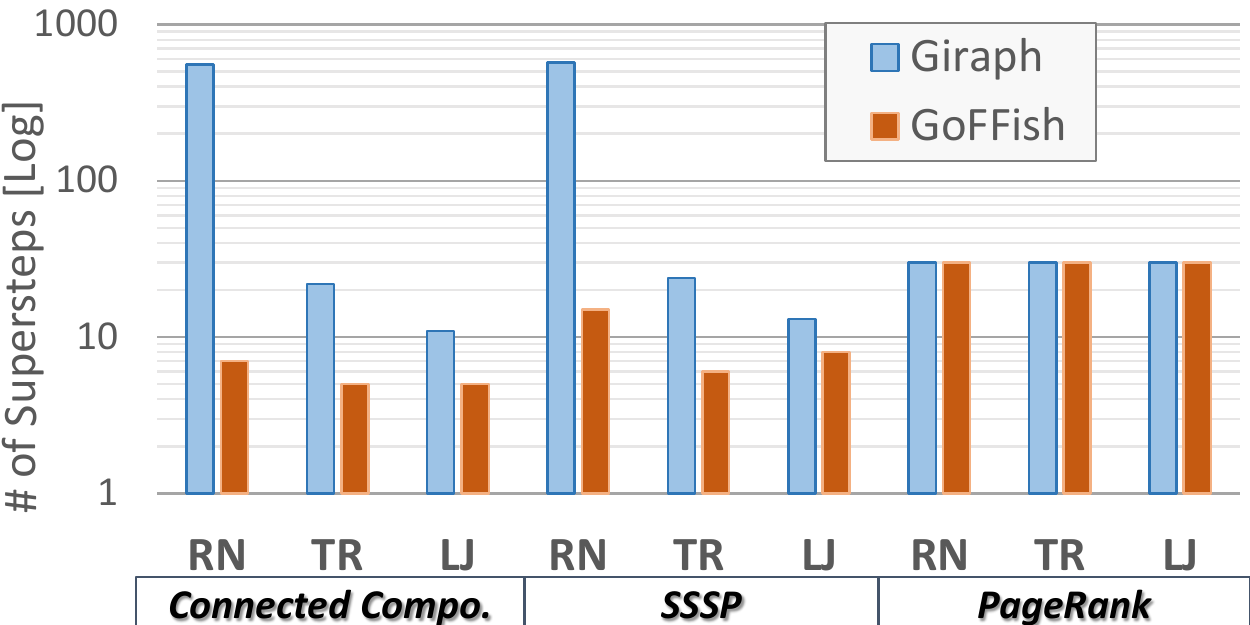}
	\label{fig:supersteps}
}

\caption{Comparison of GoFFish and Giraph for all Graph Algorithms and Datasets. Plots are in log
  scale.}
\vspace{-0.2in}
\end{figure*}

We compare the end-to-end time (makespan) for executing an algorithm on GoFFish and on Giraph which includes two
key components: the time to load the data from storage and the time to run the sub-graph/vertex
centric computation.  Fig.~\ref{fig:total} summarizes the total execution time for
various algorithms and datasets for both platforms. Also shown in Fig.~\ref{fig:supersteps} is the
number of supersteps taken to complete the algorithm for each combination. Lastly,
Fig.~\ref{fig:dataload} highlights the data loading time per graph on both platforms -- the time to
initially fetch the graphs from disk to memory objects, which does not change across algorithms if
the same attributes are loaded.

One key observation is that GoFFish's makespan is smaller than Giraph for all combination but two,
PageRank and SSSP on LJ. The performance advantage ranges from being $81\times$ faster for Connected
Components on RN to $11\%$ faster for PageRank on TR. These favorable numbers span across graphs and
algorithms, but result from abstraction, design and layout choices; likewise for the two slower
runs. In some, Giraph's data loading time from HDFS dominates (TR), in others, Gopher's sub-graph
centric algorithmic advantage significantly reduces the number of supersteps (RN for Connected
Components and SSSP), while for a few, the Gopher's compute time over sub-graphs dominates (PageRank
on LJ). These are discussed further.


\subsection{Connected Components}
Connected Components for GoFFish performs significantly better than Giraph for all three data sets
-- from $1.4\times$ to $81\times$. This is largely due to the algorithmic advantage that is offered
by a sub-graph centric model. Fig.~\ref{fig:supersteps} shows the number of supersteps is much
smaller for Gopher compared to Giraph, taking between 5 (TR, LJ) and 7 (RN) supersteps for Connected
Components while Giraph takes between 11 (LJ) and 554 (RN). This reduction in the number of
supersteps is the key reason that we see the makespan reduce sharply for GoFFish, and the superstep
time in itself is dominated by the synchronization overhead since the actual sub-graph/vertex
computation done for Connected Components is negligible.

As posited before, the number of supersteps for a traversal algorithm like Connected Components is
bounded by the graph diameter. For sub-graph centric algorithms, this is the graph diameter when
considering each sub-graph as a meta vertex. If we remove the data loading time
(Fig.~\ref{fig:dataload}) from the makespan to look at just at the ratio of compute times between Giraph
and Gopher for Connected Components, we observe that this compute improvement ratio is highly
correlated ($R^{2}=0.9999$) with the vertex-based diameter of the graph (Table~\ref{table:datasets}),
\emph{i.e., larger the vertex-based graph diameter, greater the opportunity to reduce sub-graph-based diameter,
lesser the number of supersteps for traversal algorithms, and better that Gopher performs.}

The TR graph has a smaller vertex-based diameter and yet its makespan performance for Connected
Components on Gopher is $21\times$ better than Giraph. This improvement is primarily due to much
faster data loading in GoFS ($38 secs$) compared to HDFS ($798 secs$) for TR
(Fig.~\ref{fig:dataload}). The data layout of GoFS was designed for Gopher and ensures that 
network transfer is not needed when loading the sub-graphs initially. However, Giraph uses
HDFS, which by default balances the vertices across partitions. In this case, there is one vertex
with a large edge degree (O(millions)) 
that takes punitively long to load into memory objects. The compute time
itself is just $1 sec$ and $23 secs$ for Gopher and Giraph respectively.

For LJ, we do not see a significant improvement for GoFFish over Giraph, with a $3 sec$ improvement
in load time and $18 sec$ advantage in compute time. LJ is a dense, highly connected graph with a
small vertex-based diameter. It does not offer a large reduction in the number of supersteps from
Giraph to Gopher ($11 \rightarrow 5$), nor does HDFS under-perform for this graph. In fact, as other
results show, LJ is the kind of graph that offers limited benefits for a sub-graph centric approach.

\subsection{SSSP}
SSSP for GoFFish out-performs Giraph by $78\times$ and $10\times$ for RN and TR, respectively, while
it is marginally slower for LJ by $5 secs$, compared to Giraph's $60 secs$. Gopher takes
similar number of supersteps for both Connected Components and SSSP, but its computational complexity per
superstep is different for the two algorithms: $O(v)$ vs. $O(e \cdot log(v))$. 
As a result, the time spent on
running \textsc{Djikstras} on the sub-graphs for SSSP is comparable to the time spent on I/O (for
Connected Components, compute time was negligible). This fraction is higher for graphs with higher edge
density, such as LJ. E.g., we spend $4 secs$ for SSSP compute on RN ($\sim2M$ vertices, $2.7M$
edges) while we spend $42 secs$ on LJ ($\sim4.8M$ vertices, $68M$ edges). 

This heightened computational complexity of SSSP for Gopher impacts its relative performance with Giraph too.
Giraph's vertex centric \texttt{Compute} performs a simple check and forward operation, with
constant time complexity. As a result, we see that for LJ with high edge density, the sub-graph
centric model cumulatively takes as long as Giraph for the compute time, at $43 secs$ against Giraph's $42
secs$, and both spend about the same time on initial data loading. As noted before, LJ is a small
world network with small diameter, and the reduction in the number of supersteps between Giraph and
Gopher is modest (Fig.~\ref{fig:supersteps}). Hence, for LJ, the advantage of fewer supersteps
is offset by the higher cost per superstep.

Gopher does outperform Giraph on compute time, data loading time and number of supersteps for RN
and TR, taking $4 secs$ and $53 secs$ for computation against Giraph's $1,182 secs$ and $102
secs$. Both these graphs have sparse edges densities (smaller impact on compute time) and RN has a
wider vertex based diameter too (fewer supersteps).

\subsection{PageRank}
PageRank is a canonical web graph algorithm and well suited for a vertex centric model. All vertices
can operate independently, and converge iteratively to their rank value. The computation per
superstep is uniform as is the messages exchanged in each. The vertex parallel algorithmic
simplicity of PageRank gives it good performance on Giraph. Gopher implements the classic PageRank,
simulating one iteration of vertex rank updates within a sub-graph per superstep, and running for
the same $30$ supersteps as Giraph. As a result, it does not benefit from reducing the number of
supersteps. 

As shown in Fig.~\ref{fig:total}, Gopher shows the least improvement against Giraph for PageRank. It
is $4\times$ and $1.5\times$ faster for RN and TR. However, it is $2.6\times$ \emph{slower} than
Giraph for LJ. In fact, if we ignore the initial graph loading time (Fig.~\ref{fig:dataload}), both
TR and LJ are slower in Gopher by $2.6\times$. The cause for this can be traced to under-utilization
of cores within a machine and machines across the cluster due to the variable times taken by
different sub-graphs to complete.

\begin{figure*}[ht]
\centering
\subfigure[Box-and-whiskers plot for sub-graph time distribution for TR]{
\includegraphics[width=0.8\textwidth]{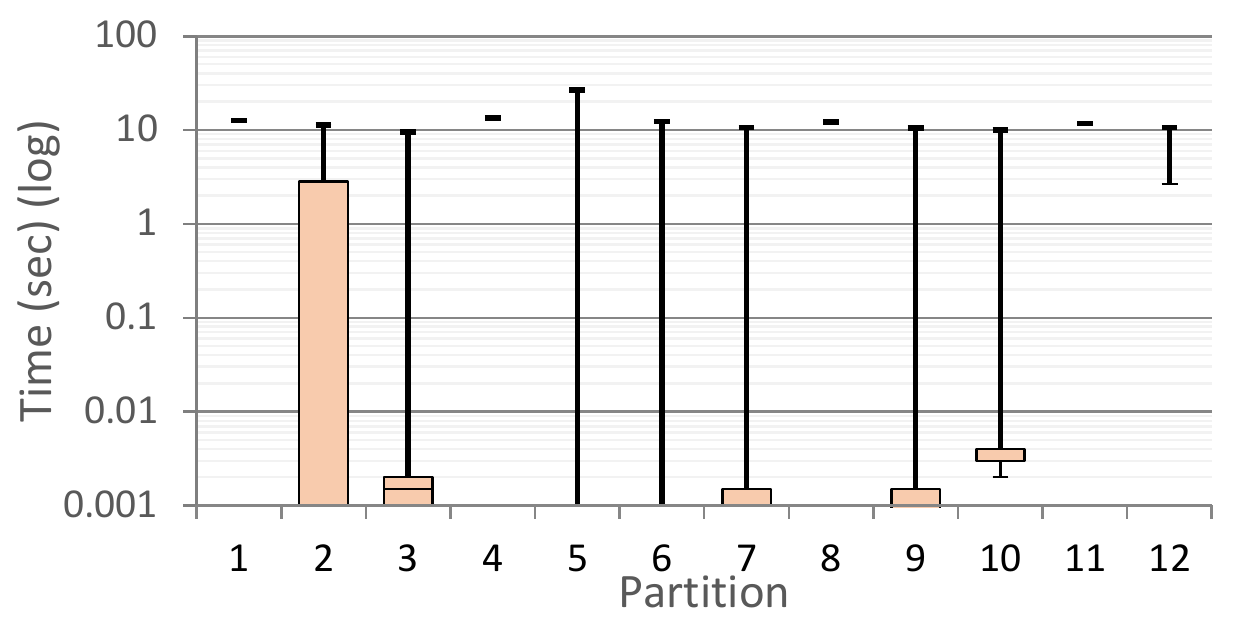}
\label{fig:ak-pr-super}
}\\
\subfigure[Box-and-whiskers plot for sub-graph time distribution for LJ]{
\includegraphics[width=0.8\textwidth]{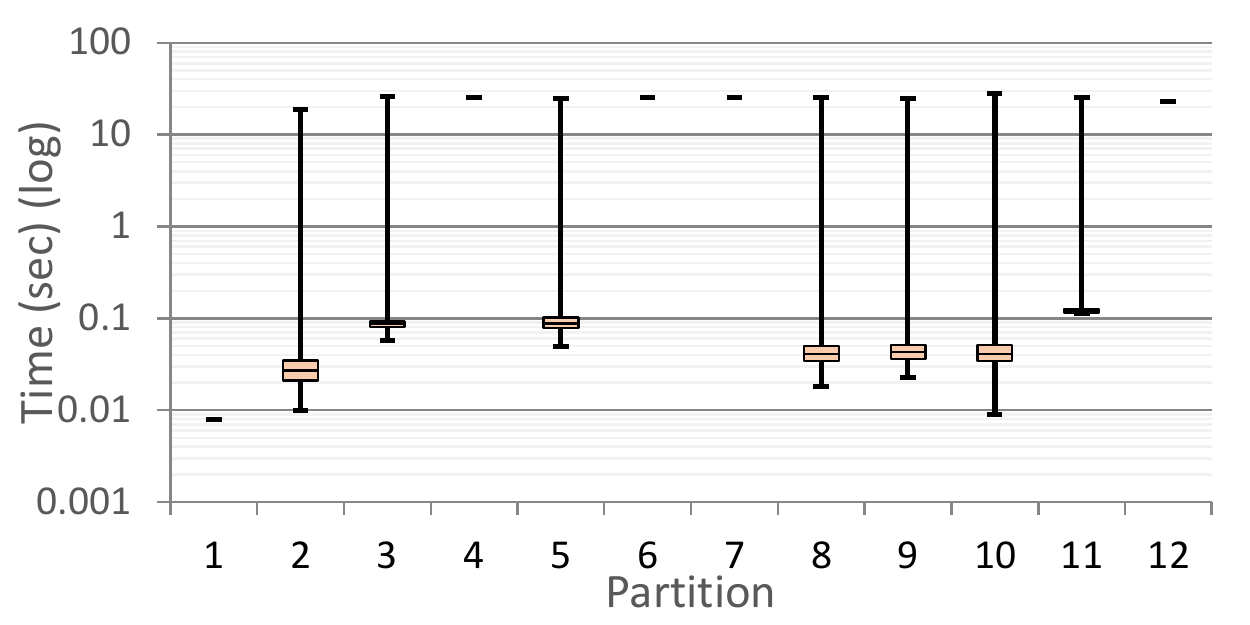}
\label{fig:lj-pr-super}
}
\caption{Distribution of sub-graph compute times within each partition for first superstep of
  PageRank}
\vspace{-0.2in}
\end{figure*}

In TR, there is one partition 5 that is a straggler in Gopher, taking $\sim 2.4
\times$ longer to complete than the next slowest, partition 3
(Fig.~\ref{fig:ak-pr-super}). This means that workers in the 11 other machines are idle for over
$58\%$ of the superstep duration as the worker in partition 5 completes. This under-utilization of
machines accumulates over the 30 superstep.

For LJ, the problem is subtlety different. While many of the partitions complete their superstep
within a narrow range of $23 - 26 secs$ (Fig.~\ref{fig:lj-pr-super}), these are due to single large
sub-graphs in each partition, i.e., these sub-graphs are stragglers within a machine causing
under-utilization of the three other cores. In fact, across partitions, the second slowest sub-graph
finishes within $0.1 secs$, causing $75\%$ of the cores to be idle for most of the superstep. This
again is cumulative across the 30 supersteps. The RN does not suffer from such imbalances; its plot
is omitted for brevity.

Giraph, on the other hand, has uniform vertex distribution across machines and each worker takes
almost the same time to complete a superstep while fully using the multi-core support through fine
grained vertex level parallelism. This highlights the deficiencies of the default partitioning model
used by GoFS that reduces edge cuts and balances the number of vertices per machine, \emph{without
  considering the number of sub-graphs that are present per partition, and their sizes}. 

\section{Conclusions}
\label{sec:conclusion}

We introduce a sub-graph centric programming abstraction for large scale graph
analytics on distributed systems. This model combines the scalability of vertex centric
programming with the flexibility of using shared-memory algorithms at the sub-graph
level. The GoFFish framework offers Gopher, a distributed execution runtime for this abstraction,
co-designed with GoFS, a distributed sub-graph aware file system that pre-partitions and
stores graphs for data-local execution. Our abstraction offers algorithmic
benefits while our architecture offers design optimizations.

We empirically showed that GoFFish performs significantly better than Apache Giraph, the popular
implementation of the vertex centric programming model for several common graph algorithms. These
performance gains are achieved due to both the partitioned graph storage and sub-graph based
retrieval from GoFS, and a significant reduction in the number of supersteps that helps us complete
faster, with lower framework overhead, that is attributable to the programming model. This offers a
high compute to communication ratio. While data parallelism is not at the vertex level, we offer
parallelism across sub-graphs and the option to use fast, shared-memory kernels within a sub-graph.


We do recognize some short comings, that offer further research opportunities into this promising
model. Sub-graph centric algorithms are vulnerable to imbalances in number of sub-graphs per
partition as well as non-uniformity in their sizes. This causes stragglers. Better partitioning
algorithms to balance the sub-graphs can help. The framework is also susceptible to small-world
network graphs with high edge degrees that have high sub-graph level computational complexity, such
as Live Journal. Trivial mapping of convergence algorithms like PageRank need to be replaced by
algorithms such as BlockRank that effectively leverage the abstraction. Our early software prototype
offers further opportunities for design and engineering optimizations.



\section*{Acknowledgment }
This work was supported by the DARPA XDATA program. The views and opinions of authors expressed herein
do not necessarily state or reflect those of the United States Government or any agency thereof. We
thank H. Chu, N. Ma, D. Tong, M. Frincu and C. Chelmis from USC, and
other XDATA performers for their help in preparing this manuscript.
\bibliographystyle{IEEEtran}

\bibliography{main}

\end{document}